\documentclass[conference]{IEEEtran}
\IEEEoverridecommandlockouts
\usepackage{cite}
\usepackage{amsmath,amssymb,amsfonts}
\usepackage{algorithmic}
\usepackage{graphicx}
\usepackage{textcomp}
\usepackage{xcolor}
\usepackage{authblk}
\usepackage[]{footmisc}
\usepackage{url}
\usepackage[numbers]{natbib}
\usepackage{multirow}
\usepackage{subfigure}
\usepackage{tabularx,booktabs}
\usepackage{color}
\usepackage{marvosym}
\usepackage{multicol}
\def\BibTeX{{\rm B\kern-.05em{\sc i\kern-.025em b}\kern-.08em
    T\kern-.1667em\lower.7ex\hbox{E}\kern-.125emX}}
    
\begin{document}

\title{Micro-Behavior Encoding for Session-based Recommendation}
\author[1]{Jiahao~Yuan}
\author[1]{\Letter Wendi~Ji}
\author[2,3]{Dell Zhang\textsuperscript{\textsection}}
\author[1]{Jinwei Pan}
\author[1]{Xiaoling Wang}

\affil[1]{Shanghai Key Laboratory of Trustworthy Computing, East China Normal University, Shanghai, China}
\affil[2]{Birkbeck, University of London, UK}
\affil[3]{ByteDance AI Lab, London, UK}
\affil[ ]{\{jhyuan, jwpan\}@stu.ecnu.edu.cn, \{wdji, xlwang\}@cs.ecnu.edu.cn, dell.z@ieee.org}

\maketitle
\begingroup\renewcommand\thefootnote{\textsection}
\footnotetext{Dell Zhang is currently on leave from Birkbeck, University of London and working full-time for ByteDance AI Lab, London, UK.}
\endgroup

\begin{abstract}
Session-based Recommendation (SR) aims to predict the next item for recommendation based on previously recorded sessions of user interaction. 
The majority of existing approaches to SR focus on modeling the transition patterns of \emph{items}. 
In such models, the so-called \emph{micro-behaviors} describing how the user locates an item and carries out various activities on it (e.g., \textit{click}, \textit{add-to-cart}, and \textit{read-comments}), are simply ignored. 
A few recent studies have tried to incorporate the \emph{sequential} patterns of micro-behaviors into SR models. 
However, those sequential models still cannot effectively capture all the inherent interdependencies between micro-behavior operations. 
In this work, we aim to investigate the effects of the micro-behavior information in SR systematically. 
Specifically, we identify two different patterns of micro-behaviors: ``sequential patterns'' and ``dyadic relational patterns''. 
To build a unified model of user micro-behaviors, we first devise a multigraph to aggregate the sequential patterns from different items via a graph neural network, and then utilize an extended self-attention network to exploit the pair-wise \emph{relational} patterns of micro-behaviors. 
Extensive experiments on three public real-world datasets show the superiority of the proposed approach over the state-of-the-art baselines and confirm the usefulness of these two different micro-behavior patterns for SR.
\end{abstract}


\section{Introduction}
Recommender systems are a subclass of \emph{information retrieval} applications where the information need is typically represented by not an explicit query but a user's profile and context~\cite{garcia-molinaInformationSeekingConvergence2011}. 
The core functionality of such online systems is to predict the rating or preference the users would give to different items by analyzing their past behaviors. 
In many commercial websites that employ recommender systems, the sequence of a user's interactions can be naturally segmented into a number of sessions each of which occurs within a certain period of time and reflects the user's interest at that moment.
Due to user privacy concerns, real-life recommender systems are often unable to identify each user and track their long-term interests. 
However, users usually have a very specific and clear short-term intention when they visit an e-commerce or other online service websites~\cite{jannach2015adaptation}. 
Therefore, \emph{Session-based Recommendation} (SR) which aims to predict the next item for a user in the current session, has recently attracted a lot of attention from Internet companies~\cite{quadrana2018sequence}.

Most existing methods for SR focus on modeling the transition patterns of \emph{items} within a session through techniques like Markov chain~\cite{rendle2010factorizing}, Recurrent Neural Network (RNN)~\cite{li2017neural}, attention mechanism~\cite{liu2018stamp}, and Graph Neural Network (GNN)~\cite{wu2019session, xu2019graph, pan2020star}. 
Although the field of SR has witnessed significant progress in the last few years, it is still facing some challenging problems. 
Firstly, those methods only analyze the sequence of items in a session but discard the detailed operations carried out by the user on each item. 
However, the sequence of items, also known as \emph{macro-behaviors}~\cite{zhou2018micro}, could not paint the complete picture.  
Compared with the coarse-grained sequence of items, the fine-grained sequence of different operations performed on each item should be able to reflect the user's intentions and preferences more precisely. 
In this paper, we refer to such operations with respect to a particular item within a session as \emph{micro-behaviors}, and try to make effective use of them for SR.
Furthermore, although GNN-based methods have achieved exciting results and offered a promising direction for modeling the transition patterns of macro items, they cannot incorporate multiple micro-operations of the same item with their graph construction and information aggregation methods.
Secondly, to the best of our knowledge, a few recently emerged studies attempting to incorporate micro-behaviors into SR all just model the sequential pattern of micro-behaviors with RNN~\cite{zhou2018micro, gu2020hierarchical, meng2020incorporating}. 
However, micro-behaviors are often correlated with each other, and it would be very difficult for RNN-based sequential models to capture such interdependencies beyond the immediate predecessor/successor relations. 
In our opinion, it would be beneficial to analyze the \emph{relational} patterns of micro-behaviors in addition to the \emph{sequential} patterns. 

\begin{figure*}
	\centering
 	\includegraphics[width=0.8\textwidth]{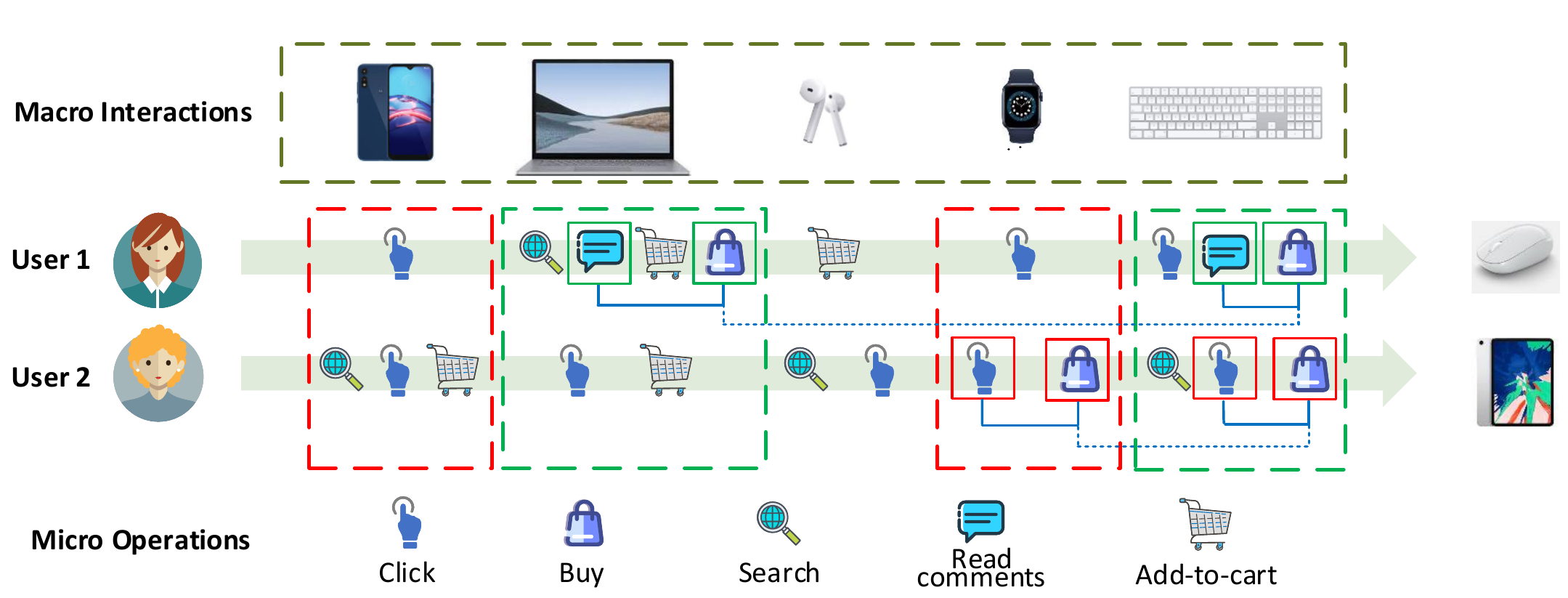}
 	\caption{Given the same item sequence with different operations, our model can make different predictions for the next item.}
 	\label{fig:background}
\end{figure*}

Fig.~\ref{fig:background} shows an example with two imaginary users. 
At the level of macro-behaviors, user 1 and user 2 are not distinguishable as they have exactly the same session: the sequence of items are iPhone, MacBook, AirPods, Apple-Watch and Magic-Keyboard. 
By contrast, at the level of micro-behaviors, user 1 and user 2 look quite different. 
Firstly, the former was probably buying computer hardware for her work since she bought MacBook and Magic-Keyboard, while the latter might be only interested in the products made by Apple Inc. 
Thus, the micro-operations in different items can help to distinguish the intentions of the user within a session. 
Secondly, user 1 tends to read the customer comments before making a purchase, while user 2 goes straightaway to place order. 
Such micro-behavioral differences would not be easily captured by a sequential model, but they could be explicitly modeled using dyadic relational patterns $\langle\texttt{read-comments}, \texttt{purchase}\rangle$ and $\langle\texttt{click}, \texttt{purchase}\rangle$.   
In other words, dyadic relational patterns may have more discriminative power than sequential patterns for the purpose of SR. 

According to the above observations, the sequential patterns are reflected in the successive micro-operations of the single item, which relate to users' preference of the item. 
The dyadic relational patterns, on the other hand, convey the pair-wise semantics of micro-behaviors which contain different meanings about the items. 
In this paper, we propose a novel framework to \underline{E}ncode \underline{M}icro-\underline{B}ehaviors in \underline{S}ession-based \underline{R}ecommendation (named \textbf{EMBSR}\footnote{The implementation has been released at \url{https://github.com/SamHaoYuan/EMBSR}}) to jointly learn those two different patterns.
Specifically, EMBSR captures the user's intention and preference in an interaction session by integrating the sequential patterns with the (dyadic) relational patterns of micro-behaviors, and thus achieves significant performance gains in recommendation accuracy. 
Firstly, we propose a new encoding scheme for converting a session into a multigraph in order to incorporate micro-behavior in GNNs. The graph construction is the premise of the proposed framework to aggregate the information of micro-behaviors since a macro item may correspond to multiple successive micro-operations.   
Secondly, we employ Gated Recurrent Unit (GRU)~\cite{cho2014learning} to describe the sequential pattern for each single item, which has aggregated micro-behavior information in iterations of GNNs. Next, inspired by self-attention with relative position representations~\cite{shaw2018self}, we develop an operation-aware self-attention mechanism which incorporates the dyadic operation representation into the self-attention network for exploiting the pair-wise semantics of micro-behaviors. 
Finally, we combine such different representations of the user's interest with the fusion gating mechanism to predict the next item. 
 
The main contributions of this paper are as follows. 
\begin{itemize}
    \item We consider a user's interaction session as a fine-grained sequence of micro-behaviors and discover two different types of patterns (sequential and relational) for the modeling of micro-behaviors in SR.
    \item We design a novel encoding scheme that transforms a session into a directed multigraph and employ a novel aggregation stage to incorporate the sequential patterns in the graph using GRUs.
    \item We propose a novel operation-aware self-attention mechanism to encode the dyadic relations of micro-behaviors, and combine the valuable information embedded in a session's sequential and relational patterns to enhance the prediction of next item.
    \item We conduct extensive experiments on three public real-world datasets to demonstrate that our proposed model can outperform all the state-of-the-art baselines for SR significantly.
\end{itemize}

\section{Preliminaries}

In this section, we introduce the preliminary knowledge about GNNs and give its message passing paradigm, and then we formulate the problem to address in this paper. 
\subsection{Graph Neural Network}
GNN generalizes traditional neural networks to graph for capturing structural information. Take advantage of the powerful modeling capabilities for nodes and edges, the task of SR is formulated as the graph classification problem.

Let $\mathcal{G}(\mathcal{V}, \mathcal{E})$ be a graph with nodes $\mathcal{V}$ and edges $\mathcal{E}$, $(u, e, v)$ denote an edge $e \in \mathcal{E}$ where $u \in \mathcal{V}$ is the source node and $v \in \mathcal{V}$ is the target node. Here we use a symbol $e$ to distinguish different edge from $u \to v$. Let $x_v \in \mathbb{R}^{d_1}$, $x_u \in \mathbb{R}^{d_1}$ denote the representation of two nodes respectively, $w_e \in \mathbb{R}^{d_2}$ denote the edge feature.  Generally, the process of GNNs to learn node embeddings in the $l+1$-th layer can be formulated as follows:
$$
\begin{aligned}
  m_e^{l+1} &= f_m (x_v^l, x_u^l, w_e^l), (u,e,v)\in \mathcal{E}  \\
  a_e^{l+1} &= f_a(\{m_e^{l+1}: (u, e, v) \in \mathcal{E}\}) \\
  x_v^{l+1} &= f_u(x_v^l, a_e^{l+1})
\end{aligned}
$$
In the above equation, $f_m$ is the message function defined on each edge to generate a message by using the edge feature and its endpoint feature, and$m_e^{l+1}$ denotes the message information of this edge. $f_a$ is an aggregation function that aggregates neighbors’ information of the node along edges, $a_e^{l+1}$ denotes the aggregated information of all edges of this node. Different information can be considered here based on the incoming or the outgoing edges of the node.  Lastly, an update function $f_u$ is applied to generate a new representation of the node, and $x_v^{l+1}$ denotes the novel representation of the node $v$. After information propagates for multiple layers, we can get a new representation of each node (or edge sometimes) and use it in the downstream task.

Note that all notations in this subsection are only used here to explain the preliminary knowledge of GNN and may be inconsistent with the following expression. 

\subsection{Problem Statement}
Let $V$ and $O$ denote the set of distinct items involved in all sessions and the set of distinct operations a user can perform, respectively. 
Given the sequence of micro-behaviors in a session, $S_t = \{s_1,s_2,...,s_t\}$, where $s_i$ is the $i$-th micro-behavior of the user and $t$ is the maximum length of the session. Specifically, $s_i = (v_i, o_i)$ is a tuple that combines an item and its corresponding operations. 

To capture the fine-grained preference for the item, we first merge the successive micro-behaviors that characterize the same item to get the chronological sequence of macro-items $S_t^v = \{v^1, v^2,...,v^n \}$ and its corresponding micro-operation sequence $S_t^o=\{o^1, o^2,...,o^n\}$, where $n \leq t$ denote the length of the macro-item sequence. For each item $v^i \in S_t^v$, its operation sequence is $o^i =\{o^i_1,o^i_2,...,o^i_k\}$, because one item usually has multiple corresponding operations. That is to say, we treat the successive micro-behaviors of the same item as a whole. The goal of the proposed model is to predict the next macro-item $v^{n+1}$. Noting that we do not directly predict the next micro-behavior $s_{t+1}$ or its item $v_{t+1}$, since the last macro-item may also have multiple micro-behaviors. In other words, there is a high probability of $v_{t} = v_{t+1}$, which leads to information leakage.

To achieve this goal, the proposed model learns to generate a session representation $m$ based on the given session, and uses it to calculate a score $\hat{y_i}$ for each item $v\in V$. Finally, the items of the top-$K$ scores will be returned to the user as recommendations. For clarity, Tab.~\ref{tab:notations} summarizes the main notations and their meanings used in this paper.

\setcounter{table}{0}
\begin{table}[tp]
    \centering
    \caption{summary of notations}
    \label{tab:notations}
    \begin{tabular}{l|p{0.3\textwidth}}
    \toprule
    \textbf{Notations} & \textbf{Descriptions} \\
    \midrule
        $V$, $O$ & The item set and operation set \\
         \midrule
         $S_t$, $S_t^v$, $S_t^o$ & the given session,  the corresponding macro-item sequence,  the corresponding micro-operation sequence  \\ 
         \midrule
         $s_i$, $v_i$, $o_i$ & the $i$-th micro-behavior, item, and operation of the given session, where $s_i = (v_i, o_i)$  \\ 
         \midrule 
         $v^i$, $o^i$, $o^i_j$& the $i$-th macro item, the micro-operation sequence of $v^i$, the $j$-th micro-operation in $o^i$\\ 
         \midrule
         $\mathcal{G}_t, \mathcal{V}_t, \mathcal{E}_t$& the directed multigraph, the node set, the edge set\\
        
         \midrule
         $M^V$, $M^O$& embedding metrics of item and operation  \\
         \midrule
         $M^P$, $M^R$& embedding metrics of position and dyadic relation\\
         \midrule
         $S_t^u$, $u_i$ & the distinct item set of $S_t$, the distinct item in $S_t$ (it is also the node in the multigraph)\\ 
         \midrule
         $e_{u_i}$, ${e_{o_i}}$& the embeddings of  $u_i$ and $o_i$   \\
         \midrule
         $v_s$, $e_{u_s}$ & the star node and its embedding\\
          \midrule
          $\tilde{h}^i$, $\tilde{h}^i_j$, $\tilde{h}_o$ & the hidden state of the operation $o^i_j$,  the sequential encoding of $o^i$,  the sequential pattern encoding of the micro-operation in the given session\\
         \midrule
         $m^{l}_{i+}$, $m^{l}_{i-}$, $m^{l}_{i}$ & the message for incoming edges, outgoing edges and all edges of the node $u_i$ at layer $l$ \\
         \midrule
         $E_{in}(i)$, $E_{out}(i)$ & the set of incoming edges and outgoing edges of the node $u_i$ \\
         \midrule
         $h^i$, $h^f$ & the hidden layer of GNN, the final representation of satellite nodes \\
         \midrule
         $M_{s_t}$, $r_{ij}$, $e_{r_{ij}}$& the relation matrix of the session $S_t$,  the dyadic relation index between $o_i$ and $o_j$,  the embedding of the dyadic relation $r_{ij}$ \\
         \midrule
         $X_t$, $x_i$, $x_s$ & the input embeddings for the operation-aware self-attention mechanism, the embedding of the tuple $s_i$,  the embedding of the star node \\
         \midrule
         $e_{p_i}$, $e_{ij}$& the embedding of the $i$-th position,  the intermediate value to measure the correlation between $x_i$ and $x_j$\\
         \midrule
         $z_s$, $m$ & the output of the operation-aware self-attention mechanism,  the final output of the proposed approach \\
         \bottomrule
    \end{tabular}

\end{table}

\section{Related Work}

\subsection{Session-based Recommendation}

Since the long-term user profile is unknown, the task of SR is limited to the context within the sessions. Hence, for conventional methods, simple matrix factorization~\cite{mnih2008probabilistic, koren2015advances} and Item-KNN~\cite{sarwar2001item} without considering the order of items are not suitable for SR. Thus, FPMC \cite{rendle2010factorizing} used tensor factorization that combines Markov chain to simulate the sequential behavior between two adjacent clicks. \citeauthor{kamehkhosh2017comparison} \cite{kamehkhosh2017comparison} combined Markov chain with association rules. Recently, methods based on nearest-neighbors obtain competitive performance. SKNN~\cite{jannach2017recurrent} is a session-based k-nearest-neighbors approach, which considers the sessions that contain any item of the current session as neighbors. STAN \cite{garg2019sequence} is an extension version that additionally considers other factors. However, this kind of approach is generally limited to represent complex dependencies.

Neural network methods are proposed to capture the dynamic preferences from user's sequential behaviors, which have gained momentum in SR. GRU4Rec \cite{Hidasi2016} first introduced GRU to model user sessions. GRU4Rec+ \cite{tan2016improved} proposed a data augmentation method and took the changes in user behavior over time into account. Moreover, NARM~\cite{li2017neural} proposed an encoder-decoder architecture based on RNN and attention mechanism. STAMP~\cite{liu2018stamp} used the attention mechanism to capture both the users' long-term and short-term interests. Bert4Rec~\cite{sun2019bert} employed the deep bidirectional self-attention to identify the correlations of items. These attention-based models effectively capture both the user's general and current interest. CSRM~\cite{Collaborative2019Sigir} and CoSAN~\cite{ijcai2020C} incorporates collaborative neighborhood information into neural SR models. 

Recently, GNNs have been applied to SR to capture the complex transition patterns. SR-GNN~\cite{wu2019session} first introduced a gated GNN into this task, which converts a session to graph-structured data. Furthermore, GC-SAN~\cite{xu2019graph} proposed a graph contextualized self-attention model, which utilizes both GNNs and the self-attention mechanism. FGNN~\cite{qiu2019rethinking} replaced the gated GNN with multi-layered weighted graph attention networks~\cite{2018graphAttention}. To capture the long-term dependencies, LESSR~\cite{chen2020handling} used an GAT layer to learn the global dependency by propagating information along long-range edges, and SGNN-HN ~\cite{pan2020star} introduced a star GNN for the problem. For global information, GCE-GNN~\cite{wang2020global} build a global graph to learn the global-level item embeddings by modeling pair-wise item-transitions over session, and DHCN~\cite{xia2021self} adopted a dual channel hypergraph convolutional network.

To summarize, a majority of approaches for SR have focused on exploring the complex transition of macro-items and neural methods, especially GNN-based methods, which show promising potential to model dependencies of items. However, they all neglect user micro-behaviors in sessions, which reveals the fine-grained preferences or attitudes of the user. Compared with these macro-behavior models, the proposed framework investigates the effects of the micro-behavior information in SR systematically and explores two different patterns of micro-behaviors.

\subsection{Micro-Behaviors in Recommendation}
Micro-behaviors are also integrated in the task of conventional recommendation (called multi-behavior-based recommendation~\cite{gao2019learning}), which aims to leverage the micro-behaviors to improve the recommendation performance on the target behavior (e.g. purchase). 

\citeauthor{quadrana2018sequence}~\cite{quadrana2018sequence} proposed behavioral factorization, which extends the collective matrix factorization~\cite{singh2008relational} to handle different behaviors in online social network (comment, re-share, and create-post). 
\citeauthor{liu2017multi}~\cite{liu2017multi} employed behavior-specific transition matrices in recurrent and time-aware Log-BiLinear \cite{mnih2007three} model to capture the properties of different types of behaviors. 
\citeauthor{wan2018item}~\cite{wan2018item} utilized tensor decomposition framework to model monotonic behavior chains where user behaviors are supposed to follow the same chain. 
\citeauthor{gao2019learning}~\cite{gao2019learning} proposed to correlate the model prediction of each behavior type in a cascaded way and trained the whole model in a multi-task manner. 
Moreover, \citeauthor{lo2016understanding}~\cite{lo2016understanding} analyzed the purchasing behavior of users to determine short-term and long-term signals in user behavior that indicate higher purchase intent. Multi-behaviors are also considered as features and extracted from user clickstreams to help predict purchase \cite{olbrich2011modeling, yehezki2017classifying}. 
Furthermore, from the perspective of learning, there are many works exploiting multi-behaviors as auxiliary action for sampling~\cite{loni2016bayesian, guo2017resolving, qiu2018bprh}.
However, these works verify the effectiveness of multi-behaviors to help model the target behavior in conventional recommendation scenarios but have not explored the complex patterns in SR.

Recently, a few works have begun to take micro-behaviors into consideration in SR. 
\citeauthor{zhou2018micro}~\cite{zhou2018micro} first adopted RNN to model micro-behaviors, and \citeauthor{gu2020hierarchical}~\cite{gu2020hierarchical} extended it to a hierarchical architecture to distinguish the differences between micro-operations and macro-items. 
These RNN-based models ignore the different dependencies between items and operations. 
To tackle this problem, MKM-SR \cite{meng2020incorporating} fed the operation sequence and the item sequence of a session into RNN and GNN, respectively. However, MKM-SR only considers the transition of macro items in GNNs and cannot incorporate the micro-behavior information in the iterative process of GNNs. It treated items and operations separately and only concatenated them for final session representations. In addition, all those methods only consider the sequential pattern of the micro-operation and are very difficult to capture the interdependencies of dyadic relational patterns. In contrast, our work devises a multigraph to aggregate the sequential patterns and then utilizes an extended self-attention network to exploit the pair-wise relational patterns of micro-behaviors. 


\section{APPROACH}
In this section, we present the proposed model in detail. We  first summarize  the  pipeline of the proposed model. Then we describe the two major components of the proposed method to encode sequential patterns and dyadic relational patterns of micro-behaviors. Lastly, we introduce a prediction layer to generate the session representation with a fusion gating mechanism. 

\subsection{Model Overview}
\begin{figure*}[t]
    \centering
    \includegraphics[width= \textwidth]{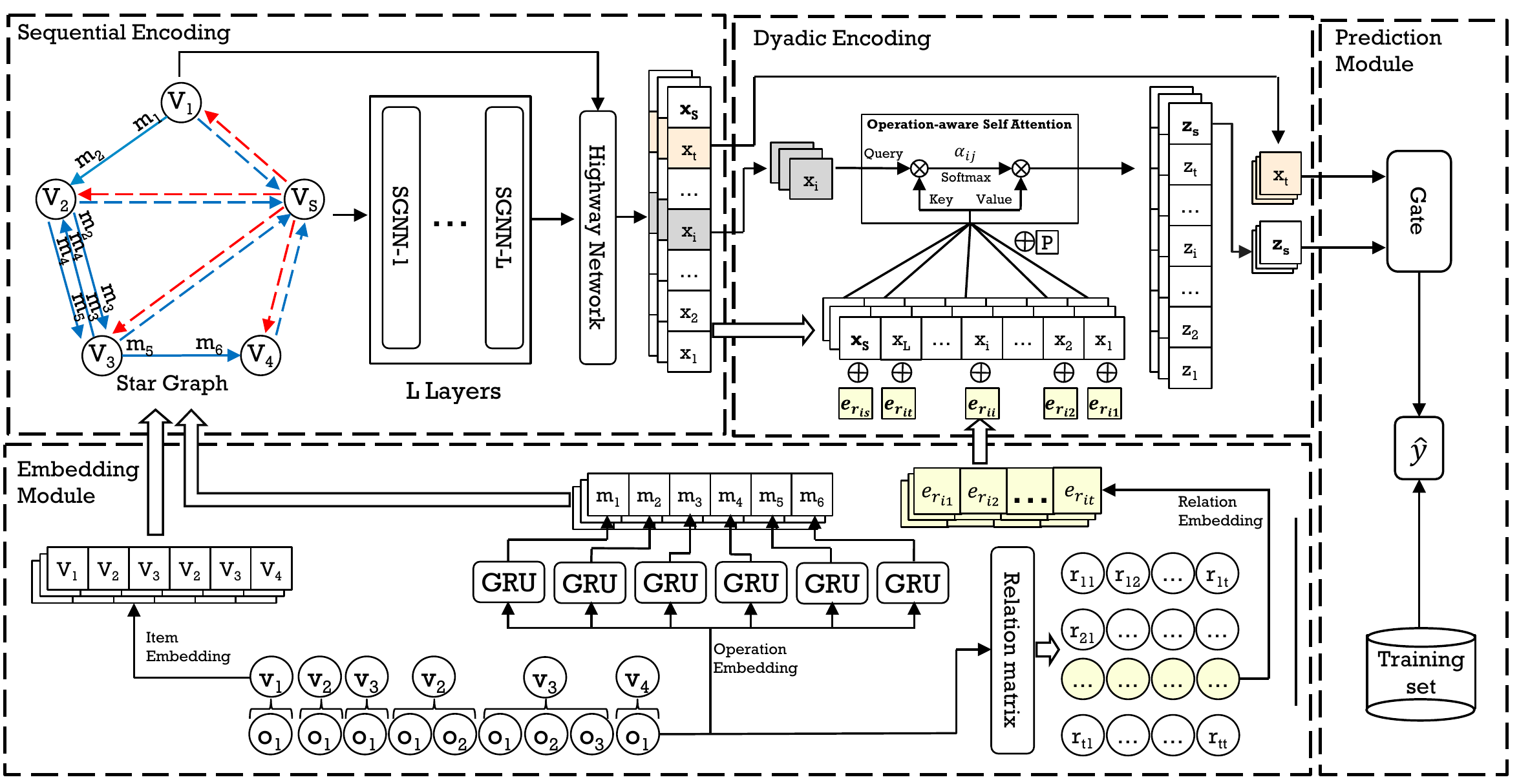}
    \caption{Overview of the proposed EMBSR model.}
    \label{Model}
\end{figure*}
The framework of the proposed model is illustrated in Fig \ref{Model}.
For an input session, we transform each micro-behavior into item embedding, operation embedding, and relation embedding. 
Then for encoding sequential patterns, we convert the macro-item sequence of the input session into a multigraph with a star node and use a GRU for the micro-operation sequence to get the feature on edges. Here, we learn a new item embedding for each macro-item by the star graph. 
For encoding dyadic relational patterns, the new item embeddings incorporating the relation embeddings are fed into an operation-aware self-attention mechanism to exploit the pair-wise semantics of micro-behaviors.
Lastly, the session is represented by combining a general preference and a recent interest in the session with a fusion gating network for generating the scores on all candidate items. 

\subsection{Encoding Sequential Patterns}

Based on the basic principle of SR~\cite{quadrana2018sequence}, the premise of obtaining a session representation is to learn each object’s embedding in the session. In the setting of this paper, the object is the micro-behaviors, including items and operations. Compared with the operation sequence of a session, the transition pattern of the item sequence is more complex and does not simply exhibit the sequential pattern~\cite{meng2020incorporating}. Thus, we adopt GNNs to model the macro-item sequence and aggregate micro-operation information by GRU in it. 

Next, we introduce how we convert the input session into a multigraph and present how to propagate information between the macro-items and micro-operations via the proposed model.

\subsubsection{Graph Construction}

\begin{figure}[!t]
    \centering
    \includegraphics[width=\columnwidth]{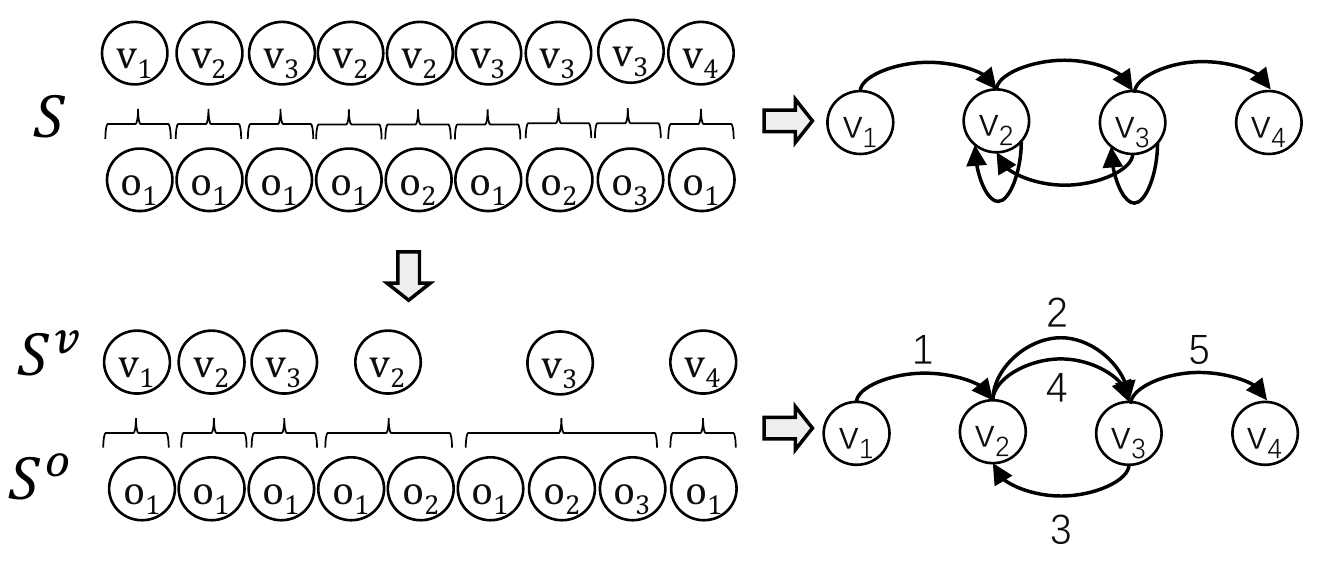}
    \caption{Different way of graph construction. We use the second way that transforming the session into a multigraph}
    \label{fig:graph}
\end{figure}

We convert the macro-item sequence to a directed multigraph that preserves the edge order. Specifically, For each macro-item sequence $S_t^v = \{v^1, v^2, ..., v^n\}$, we model it as a directed multigraph $\mathcal{G}_t = (\mathcal{V}_t, \mathcal{E}_t)$. In $\mathcal{G}_t$, each node is the distinct item in $S_t^v$ and each directed edge $(v^i, v^{i+1}) \in \mathcal{E}_t$ is the transition $v^i \to v^{i+1}$ in $S_t^v$. It is worth noting that the graph is a multigraph since there may be multiple transitions between the same items. The edges are ordered by the time of their occurrences in the session in order to distinguish the micro-operation information that their endpoints have at different times.

In Fig.~\ref{fig:graph}, we use an example to illustrate the way of converting an input session to a multigraph. For the input session $S$, it has four different items $\{v_1, v_2, v_3, v_4\} \in V $ and three different micro-operations $\{o_1, o_2, o_3\} \in O$. First of all, we merge the successive micro-behaviors with the same item to get the macro-item sequence $S^v = \{v_1, v_2, v_3, v_2, v_3, v_4\}$ and its corresponding sequence list $S^o = \{(o_1), (o_1), (o_1), (o_1, o_2), (o_1, o_2, o_3), (o_1)\}$. Then we convert $S^v$ to a multigraph and record the order by giving each edge an integer attribute to help propagate the micro-operation information of endpoints. Since the same item has a different micro-operation sequence on different positions, the multigraph ensures that the neighbor nodes pass the different message along edges based on the different micro-operation sequences.  

Inspired by \cite{pan2020star}, we add a \textit{star node} to capture the long-range information by propagating information from non-adjacent items, which is $v_s$ in Fig.~\ref{Model}. For convenience, we call other nodes from the macro-item sequence \textit{satellite node}. The star node has a bidirectional edge with each satellite node, but we update it in different way from satellite nodes.

\subsubsection{Initialization of Node Embedding}
Before passing the nodes into the proposed model, we construct an embedding matrices $M^V \in \mathbb{R}^{|V|\times d}$ for items, where $d$ is the latent dimension. We use $S_t^u = \{u_1, u_2,...,u_c\}$ to denote the distinct items set in the input session $S_t$, where $c$ is the number of distinct items in $S_t$. Next, for each distinct item $u_i \in S_t^u$ , we get its embeddings $e_{u_i} \in \mathbb{R}^d$ by the lookup of the item embedding matrix and use it as initial embedding of satellite nodes in $\mathcal{G}_t$. So the initial hidden layer $h^0$ is:
\begin{equation}
    h^0 = \{e_{u_1}, e_{u_2}, ..., e_{u_c}\}
\end{equation}
As for the star node, we apply an average pooling on the satellite nodes to get its initialization:
\begin{equation}
    e_{u_s} = \frac{1}{c} \sum_{i=1}^c e_{u_i}
\end{equation}

\subsubsection{Sequential Information of Micro-Operation}
 For each micro-operation $o^i_j \in S_t^o$, we also construct an embedding matrix $M^O \in \mathbb{R}^{|O| \times d}$ and get the micro-operation embedding $e_{o_j^i} \in \mathbb{R}^d$ by the lookup of $M^O$. So for each item $v^i \in S_t^v$, we have the representation of its micro-operation sequence $o^i$ as $\{e_{o^i_1}, e_{o^i_2}, ..., e_{o^i_k}\}$

To capture the sequential pattern of the micro-behavior, we apply RNN to the embedding of the micro-operation sequence. To avoid the problem of gradient vanishing, we use GRU, which is an improved variant of RNN. Then the $j$-th step of the GRU is:
\begin{equation}
    \tilde{h}^i_j = GRU(e_{o_j^i}, \tilde{h}^i_{j-1}; \Phi_{GRU})
\end{equation}
where $\Phi_{GRU}$ denotes all GRU parameters and $\tilde{h}^i_j$ is the hidden state of the operation $o_j^i$. For each micro-operation sequence $o^i \in S_t^o$, we use $\tilde{h}^i = \tilde{h}^i_k$ to represent the sequential information of it, where $\tilde{h}^i_k$ is the last hidden state of GRU. Thus, we obtain the learned embeddings as 
\begin{equation}
    \tilde{h}_o = \{\tilde{h}^1, \tilde{h}^2,...,\tilde{h}^n\}
\end{equation}
where $\tilde{h}_o \in \mathbb{R}^{n \times d}$ denote the whole sequential pattern of the micro-operation in the input session. Since the number of $\tilde{h}_o$ is the same as the number of macro-item sequence, we can match different sequential information for each node based on the order of the edges.

\subsubsection{Aggregation Stage}
The aggregation state includes two processes. The first process is to generate a message by the message function for each edge. The other process is to aggregate neighbors' information of each node along its edges by aggregation function. Our main idea for the information propagation is to consider the micro-operation influence on user's preference of items. Thus, the same node will pass the different messages along the different edges based on its micro-operation sequence in that position. So the whole process for each satellite node can be formulated as follow:
\begin{equation}
\begin{aligned}
m^{l+1}_{i+} = &f_m^+(\{e^l_{u_j}, \tilde{h}_j: (u_j, u_i) \in E_{in}(i)\})  \\m^{l+1}_{i-} = & f_m^-(\{e^l_{u_j}, \tilde{h}_j: (u_i, u_j) \in E_{out}(i)\})
\end{aligned}
\end{equation}
where $e_{u_j}^l$ is the representation of the node $u_j$ at layer $l$, $E_{in}(i)$ and $E_{out}(i)$ denote the set of incoming edges and outgoing edges of the node $u_i$ respectively. The message functions $f_m^+$ and $f_m^-$ are used to compute the message to be propagated from the neighbor node along the incoming and outgoing edges, which are defined as:
\begin{equation}
\begin{aligned}
f_m^+(e_{u_j}, \tilde{h}_j) = W^+_m([e_{u_j};\tilde{h}_j]) + b^+_m \\
f_m^-(e_{u_j}, \tilde{h}_j) = W^-_m([e_{u_j};\tilde{h}_j]) + b^-_m 
\end{aligned}
\end{equation}
where $W^+_m, W^-_m \in \mathcal{R}^{2d \times d}$ and $b^+_m, b^-_m \in \mathcal{R}^d$ are learnable parameters, $[\cdot]$ denote the concatenation operation. Hence, we can obtain $m^{l+1}_{i} \in \mathcal{R}^{|\mathcal{E}(i)| \times d}$ to denote the message for all edges of the node $u_i$. Noting that we do not consider the edge with the star node here to avoid destroying the structural information of the original session. After that, we aggregate messages of all edges for node $u_i$ by 
\begin{equation}
    a_{i}^{l+1} = [\sum_{k=1}^{|E_{in}(i)|}m^{l+1}_{i+,k};\sum_{k=1}^{|E_{out}(i)|}m^{l+1}_{i-, k} ]
\end{equation}
where $m^{l+1}_{i,k} \in \mathbb{R}^d$ denotes the $k$-th element in $m^{l+1}_{i} $ and $a_i^{l+1} \in \mathbb{R}^{2d}$ denotes the aggregated information of $u_i$.

\subsubsection{Update Stage}
The update stage is to update the node feature by using the aggregation information for each node. We feed $a_i^{l+1}$ and the node $u_i$'s previous embedding into the gated GNN \cite{ggcn2016} as follows:
\begin{equation}
    \begin{aligned}
    \tilde{z_i}^{l+1} &= \sigma(W_za^{l+1}_i + U_z e_{u_i}^{l}) \\
    r_i^{l+1} & = \sigma(W_ra^{l+1}_i + U_r e_{u_i}^l) \\
    \tilde{e}_{u_i}^{l+1} &= tanh(W_u a^{l+1}_i + U_u(r_i^{l+1} \odot e_{u_i}^{l})) \\
    \hat{e}_{u_i}^{l+1}& = (1-\tilde{z_i}^{l+1})\odot e_{u_i}^l + \tilde{z_i}^{l+1} \odot \tilde{e}_{u_i}^{l+1}
    \end{aligned}
\end{equation}
where $W_Z, W_r, W_u \in \mathbb{R}^{2d \times d}$ and $U_z, U_r, U_u \in \mathbb{R}^d$ are trainable parameters of the network. $\sigma (\cdot)$ denotes the $sigmoid$ function and $\odot$ is the element-wise multiplication. In addition, $\tilde{z_i}^{l+1}$ and $r_i^{l+1}$ are update gate and reset gate respectively, which controls how much information should be preserved or updated between two layers. In this way, information from satellite nodes can be propagated by the GNN.

Next, we consider the connection from the star node to explicitly capture the overall information of the session. For each satellite node, we adopt a gating network to decide how much information should be propagated from the star node and the adjacent nodes, which is as:
\begin{equation}
    \begin{aligned}
    \alpha_i^{l+1} &= \frac{(W_{q1} \hat{e}_{u_i}^{l+1})^T W_{k1}e_{u_s}^{l}}{\sqrt{d}} \\
    e_{u_i}^{l+1} &= (1-\alpha_i^{l+1})\hat{e}_{u_i}^{l+1} + \alpha_i^{l+1} e_{u_s}^{l} 
    \end{aligned}
\end{equation}
where $W_{q1}, W_{k1} \in \mathbb{R}^{d\times d}$ are both parameter matrices. This gating network determines how to selectively integrate the information from $\hat{e}_{u_i}^{l+1}$ and the former star node $e_{u_s}^l$ to generate the new representation of satellite node $u_i$.

For updating the star node, we use attention mechanism to assign different weight to the satellite nodes by regarding the star node as $query$:
\begin{equation}
\begin{aligned}
    \beta_i = softmax &(\frac{(W_{k2}e_{u_i}^{l+1})^T W_{q2}e_{u_s}^l}{\sqrt{d}}) \\
    e_{u_s}^{l+1} &= \sum_{i=1}^{l_s}\beta_i e_{u_i}^{l+1}
    \label{eq:starnode}
\end{aligned}
\end{equation}
where $W_{q2}, W_{k2} \in \mathbb{R}^{d \times d} $ are the learnable parameters, $\beta_i$ is the weight of the node $u_i$. Moreover, we apply a highway network \cite{srivastava2015highway} to selectively obtain information from the item embeddings before and after the stacked GNN layers, which is denoted as:
\begin{equation}
    \begin{aligned}
    g &= \sigma(W_g[h^0;h^{last}]) \\
    h^f &= g\odot h^0 + (1-g) \odot h^{last} 
    \end{aligned}
\end{equation}
where $W_g \in \mathbb{R}^{2d \times d}$ is the trainable parameter. By this highway network, we can obtain the final representation of satellite nodes as $h^f$ and the corresponding star node $e_{u_s}^{last}$ (denoted as $e_{u_s}$ for brevity).

\subsection{Encoding Dyadic Relational Patterns}
To encode dyadic relation patterns of micro-behaviors and generate the session representation, we should aggregate the embedding of all micro-behaviors in the input session. In the above GNNs, we have already integrated the sequential pattern of the micro-operation into the representation of the items, but the relational pattern of the micro-operation has still been neglected. In this subsection, we introduce a method to encode dyadic micro-operation and present a novel operation-aware self-attention mechanism. 

\subsubsection{Dyadic Micro-Operation Encoding}
In an attempt to model the pairwise relations between all the operations in the input session, we create a relation matrix of operation $M^R \in \mathbb{R}^{|O|^2 \times d}$. 
Each vector with dimension $d$ in $M^R$ represents a dyadic encoding of the operation pair. 
For example, suppose that we have 10 different micro-operations, combining them in pairs, there will be 100 different couples, e.g., $\langle\texttt{click}, \texttt{purchase}\rangle$, $\langle\texttt{click}, \texttt{read-comments}\rangle$. 
Therefore, for the operation sequence $O_t = \{o_1, o_2,...,o_t\}$ of the session $S_t$, the relation matrix of it is:
\[
M_{S_t} = 
\begin{bmatrix} 
r_{11} & r_{12} & \ldots & r_{1t} \\ 
r_{21} & r_{22} & \ldots & r_{2t} \\
\vdots & \vdots & \ddots & \vdots \\
r_{t1} & r_{t2} & \ldots & r_{tt} \\ 
\end{bmatrix}
\]
After retrieving the relation matrix $M^R$, we get the embedding $e_{r_{ij}} \in \mathbb{R}^d$ for $r_{ij} \in M_{S_t}$.

\subsubsection{Operation-Aware Self-Attention Mechanism}
To capture the dyadic information in micro-behaviors, we propose an extension to self-attention, which incorporates the pair-wise operation embedding in the sequence. Given the micro-behavior session $S_t$, we use the embedding sequence $X_t = \{x_1, x_2, ..., x_t\}$ as the input vectors, where $x_i \in \mathbb{R}^d$ is the representation for the tuple $s_i = (v_i, o_i) \in S_t$. It is calculated by:
\begin{equation}
    x_i = e_{v_i} + e_{o_i}.
    \label{input_embedding}
\end{equation}
where $e_{v_i} \in \mathbb{R}^d $ is the item embedding from the corresponding satellite nodes $h^f\in \mathbb{R}^{l \times d}$, $e_{o_i}$ is the operation embedding by lookup the matrix $M^O$. Since the star node has fused the entire session’s information in the GNNs, inspired by \cite{yuan2021dual}, we use it as the representation of the target item and suppose that it has the same micro-operation with the next item:
\begin{equation}
    x_s = e_{u_s} + e_{o_{t+1}}
\end{equation}
where $x_s \in \mathbb{R}^d$ can be used to denote the user's global preference and is concatenated on the end of the $X_t$. 

Then we create a learn-able embedding matrix $M^P \in \mathbb{R}^{L \times d}$ for positional information in the self-attention mechanism. In this layer, each output element, $z_i \in \mathbb{R}^d$, is computed by
\begin{equation}
    z_i = \sum_{j=1}^{t}\alpha_{ij}(x_j + e_{r_{ij}} + e_{p_j})
\end{equation}
where $e_{p_j} \in \mathbb{R}^d$ is the positional embedding for the position $j$ in the session. $\alpha_{ij}$ is the attention weight, which is calculated as:
\begin{equation}
      \alpha_{ij} = \frac{\exp(e_{ij})}{\sum_{k=1}^{t} \exp(e_{ik})}  
\end{equation}

Furthermore, $e_{ij}$ is the intermediate value, which is used to measure the correlation between the pair of micro-behaviors. 
It is computed as:
\begin{equation}
  e_{ij} = \frac{x_i W^Q (x_j + e_{r_{ij}} + e_{p_j})}{\sqrt{d}}
\end{equation}
where $W^Q \in \mathbb{R}^{d \times d}$ is the input projection for the query, which is used to make the representation more flexible. 
Compared with the standard self-attention mechanism, we have different query, key, and value vectors here. 
Essentially, the output has incorporated the information of dyadic operations, which can better reflect the intents and preferences of the user based on micro-behavior patterns.

Then we apply \emph{Position-wise Feed-Forward Network} to endow the model with more non-linearity,
\begin{equation}
    FFN(z_i) = \max(0, z_i W_1 + b_1) W_2 + b_2
\end{equation}
where $W_1, W_2 \in \mathbb{R}^{d \times d}$ are the weight matrices and $b_1, b_2 \in \mathbb{R}^d$ are the bias vectors. 
After that, we add residual connections, layer normalization, and dropout mechanism as in the basic self-attention model. Finally, we get the learned vector $z_s$ as the representation of the user's final global preference, which corresponds to $x_s$ in the input.

\subsection{Session Representation and Prediction}
To obtain a session representation, we take into account a user's global preference and recent interest. For the recent interest, we directly take the representation of the last micro-behavior to denote it, i.e. $x_t$ in the equation \ref{input_embedding}. Then, we concatenate these representation and apply a fusion gating network for the final representation of the proposed model, 
\begin{equation}
\begin{aligned}
    \beta &= \sigma(W_m [z_s; x_t] + b_m) \\
    m &= \beta \odot z_s + (1-\beta) \odot x_t
    \label{fusion}
\end{aligned}
\end{equation}
where $m\in \mathbb{R}^{d}$ denotes the final output of the proposed model, $W_m \in \mathbb{R}^{2d \times d}$ is the weighting matrix, $b_m \in \mathbb{R}^{d}$ is the bias vector. 
Lastly, for each item $v_i \in V$, we produce the prediction score as follows:
\begin{equation}
\begin{aligned}
\label{cos}
   \hat{m}  =  w_kL_2N&orm (m), \,\hat{v_i}  =  L_2Norm(v_i) \\
   \hat{y_i}  &= softmax(\hat{m}^T\hat{v_i})
\end{aligned}
\end{equation}
where $v_i$ is the initial embedding of the item $i$ and $\hat{y_i}$ denotes the probability of the item in the candidate item set $I$. $L_2Norm$ is the $L_2$ Normalization function and $w_k$ is the normalized weight. This weighted normalization \cite{gupta2019niser} and the Regularizing Softmax loss \cite{zheng2018ring} make the training process more stable and insensitive to hyper-parameters.

For training the model, we use cross-entropy as the optimization objective to learn the parameters:
\begin{equation}
    L = -\sum_{i} y_i \log (\hat{y_i})
\end{equation}

\section{Experiments}

\subsection{Settings}

\subsubsection{Datasets}

We conduct experiments on three publicly available real-world datasets. The first two datasets \footnote{The full datasets are available at \url{https://tinyurl.com/ybo8z4yz}.} are from a large Chinese e-commerce site JD.com~\cite{gu2020hierarchical}, and the other is from the RecSys Challenge 2019 \footnote{\url{http://www.recsyschallenge.com/2019/}}:  
\begin{itemize}
    \item \textbf{JD Datasets}: The datasets contain the user interaction sessions of online shopping in two product categories, ``Appliances'' and ``Computers'', respectively. There are \textbf{10} different types of micro-behavior operations in each of them, such as ``SearchList2Product'', ``Detail\_comments'' and ``Order''. 
    \item \textbf{Trivago Dataset}: The dataset is provided by trivago \footnote{\url{https://www.trivago.com/}}, which is a global hotel search platform focused on reshaping the way travelers search for and compare hotels. It contains user actions about the hotel and specifies the type of action that has been performed. We only use the train set of this challenge and take \textbf{6} types of micro-operations such as ``interaction item image'' and remove the operation whose reference value is not the item, such as ``filter selection'' and ``search for destination''.
\end{itemize}   
 For a fair comparison, we follow the previous work~\cite{gu2020hierarchical} to filter out the items with fewer than 50 occurrences in \textit{JD datasets}; use 70\%, 10\%, and 20\% of sessions as the training, validation, and testing set, respectively; and use the last item in each session as the ground truth for our SR predictions. In addition, we exclude the sessions consisting of only a single item from training and testing~\cite{wu2019session}. For \textit{trivago dataset}, the only difference is that we filter out the items with fewer than 5 occurrences. The statistics of these three datasets after preprocessing are shown in Tab.~\ref{dataset}. 
 

\begin{table}
\caption{Statistics of the datasets used.}
\label{dataset} 
\begin{tabular}{l|rrr}
	\toprule
	Datesets & JD-Appliances  & JD-Computers & Trivago\\
	\midrule
	\# train           &    583,255  &    577,301 &26,0877 \\
	\# validation      &     83,279  &     82,391 &37,027\\
	\# test            &    166,670  &    164,782 &74,770\\
	\hline 
	\# items           &     75,159  &     93,140 & 183,561\\
	\# micro-behavior  & 32,736,184  & 24,245,132 & 5,726,369\\
	\bottomrule
\end{tabular}
\end{table}

\subsubsection{Baselines}

To evaluate the effectiveness of our proposed model, we compare it with the following state-of-the-art methods for SR. 
Those baselines are classified into two categories: macro-behavior models that only utilize the sequences of items for SR, and micro-behavior models that also take the detailed operations on each item into consideration.

\textbf{Macro-Behavior Models:} 
\begin{itemize}
    \item \textbf{S-POP}~\cite{Hidasi2016,wang2018variational} simply recommends the most popular items in the current session. It is an improved version of the popularity-based method for SR. 
    \item \textbf{SKNN}~\cite{jannach2017recurrent} is the session-based $k$-nearest-neighbors approach, which scores an item based on the similarity between the target session and historical sessions. 
    \item \textbf{NARM}~\cite{li2017neural} applies RNN and the attention mechanism to capture the user's main purpose. 
    \item \textbf{STAMP}~\cite{liu2018stamp} is a short-term memory priority model, which combines the user's general interest and current interest reflected by the last-click to generate recommendation results
    \item \textbf{SRGNN}~\cite{wu2019session} is a session-based recommendation model that utilizes a graph neural network to learn the item and session representation
    \item \textbf{GCSAN}~\cite{xu2019graph} models sessions as directed graphs and makes recommendations with a self-attention network. 
    \item \textbf{BERT4Rec}~\cite{sun2019bert} uses deep bidirectional self-attention to represent user macro-behaviors.
    \item \textbf{SGNN-HN}~\cite{pan2020star} considers long-distance relations between items in a session by star graph neural network (SGNN) and applies highway networks to deal with the overfitting problem for SR. 
\end{itemize}

\textbf{Micro-Behavior Models:} 
\begin{itemize}
    \item \textbf{RIB}~\cite{zhou2018micro} is the first paper that incorporates micro-behaviors into SR by simply using a GRU.
    \item \textbf{HUP}~\cite{gu2020hierarchical} proposes a hierarchical RNN framework to harvest the sequential information of users' micro-behaviors. 
    \item \textbf{MKM-SR}~\cite{meng2020incorporating} is the latest SR model that employs GNN for item embedding and GRU for operation embedding; the variant which does not include the auxiliary task of knowledge embedding is used in our experiments since we don't have the knowledge graph of items.
    
\end{itemize}
 
\subsubsection{Metrics}

\begin{table*}[tb]
    \centering
    \caption{Performances (\%) of all the SR methods. The highest scores are boldfaced; the 2nd highest scores are underlined.}
    \label{tab:overall}
    \resizebox{\textwidth}{!}{
    \begin{tabular}{l|l|cccccccc|ccc|cc}
    \toprule
    Datasets& Metrics &S-POP & SKNN& NARM & STAMP & SR-GNN & GC-SAN& BERT4Rec & SGNN-HN & RIB & HUP & MKM-SR & EMBSR & Imp.\\
    \midrule  
    \multirow{6}*{Appliances} 
     & 
     H@5&31.66&25.06&30.94&30.74&32.65&30.36&31.02&\underline{34.80}&30.12&31.91&33.82&\textbf{37.34}& 7.30\%\\
     & H@10 &42.45&36.96&42.69&42.10&43.80&42.02&42.67&\underline{47.07}&40.84&43.39&45.02&\textbf{49.57}& 5.31\%\\
     & H@20 &49.56&49.30&54.74&53.98&55.32&54.02&54.30&\underline{59.36}&51.61&54.73&56.57&\textbf{61.64}& 3.84\%\\
     & M@5&17.29&13.15&17.90&18.21&19.63&17.83&16.96&\underline{21.00}&16.97&17.83&20.73&\textbf{23.58}& 12.29\%\\
    & M@10 &18.74&14.73&19.46&19.72&21.11&19.38&18.52&\underline{22.64}&18.40&19.37&22.22&\textbf{25.21}& 11.35\%\\
    & M@20 &19.25&15.59&20.30&20.55&21.91&20.22&19.33&\underline{23.49}&19.15&20.16&23.03&\textbf{26.06}& 10.94\%\\
    \hline
    \multirow{6}*{Computers} 
    & H@5&17.18&15.11&18.31&18.18&20.08&18.79&17.90&\underline{21.53}&16.93&18.87&21.00&\textbf{24.17}&12.26\%\\
    & H@10 &24.82&23.56&28.14&26.97&29.11&28.75&26.79&\underline{32.01}&24.56&27.62&30.21&\textbf{34.75}&8.56\% \\
    & H@20 &30.22&33.55&39.34&37.44&39.72&39.98&36.98&\underline{43.67}&33.58&37.49&40.86&\textbf{46.29}& 6.00\%\\
    & M@5&9.45&7.89&9.40&10.09&11.38&9.26&9.42&11.61&9.26&10.15&\underline{12.01}&\textbf{13.98}& 16.40\%\\
    & M@10 &10.47&9.01&10.70&11.26&12.57&10.58&10.59&13.00&10.27&1.31&\underline{13.23}&\textbf{15.38}& 16.25\%\\
    & M@20 &10.86&9.70&11.48&11.98&13.31&11.35&11.30&13.81&10.89&11.99&\underline{13.97}&\textbf{16.18}& 15.82\%\\
          \hline
    \multirow{6}*{Trivago} 
    &H@5&0&7.89&12.89&13.11&11.97&14.15&11.01&\underline{14.58}&9.00&10.06&12.34&\textbf{15.80}& 8.37\%\\
    &H@10 &0&14.03&18.02&17.13&15.91&20.10&15.00&\underline{20.13}&11.65&14.05&16.63&\textbf{22.95}& 14.01\%\\
    & H@20 &0&20.69&23.84&21.49&20.13&26.21&19.43&\underline{26.39}&14.39&18.65&21.45&\textbf{31.18}& 18.15\%\\
    & M@5&0&2.65&7.57&8.09&7.42&7.76&6.60&\underline{8.79}&5.69&6.00&7.58&\textbf{9.05}& 2.96\%\\
    & M@10 &0&3.47&8.25&8.62&7.94&8.55&7.13&\underline{9.53}&6.04&6.53&8.14&\textbf{10.00}& 4.93\%\\
    & M@20 &0&3.93&8.65&8.92&8.24&8.97&7.43&\underline{9.96}&6.24&6.85&8.48&\textbf{10.56}& 6.02\%\\
	\bottomrule
    \end{tabular}
    }
\end{table*}

Following previous works \cite{gu2020hierarchical, meng2020incorporating, pan2020star}, we adopt two commonly used performance measures for SR --- \emph{Hit Rate} (H) and \emph{Mean Reciprocal Rank} (M) at top $K$ --- to evaluate our proposed model EMBSR and the competing methods.
\begin{itemize}
    \item \textbf{H@K}: It is the proportion of cases that the ground truth is ranked amongst the top-K items.
    \begin{equation}
    H@k = \frac{n_{hit}}{N}
    \end{equation}

    where $N$ is the number of test sessions and $n_{hit}$ is the number of cases that the ground truth is included in the top $K$ list.
    \item \textbf{M@K}:It is the average of reciprocal ranks of the desired items, which is the evaluation of ranked results. When the rank is larger than k, the reciprocal rank is set to zero.
    \begin{equation}
            M@K = \frac{1}{N} \sum_{v'\in S_{test}} \frac{1}{Rank(v')} \\
    \end{equation}
    
\end{itemize}
where $v'$ is the ground truth and $S_{test}$ is the recommended list of the test dataset. 

\subsubsection{Hyperparameters}

The hyperparameters for all methods in comparison are tuned on the validation set via gird search. For all methods,  we use Adam as the model optimizer, tuned the learning rate in $[0.001, 0.003, 0.005, 0.008, 0.01]$ and the dropout rate in $[0, 0.1, 0.2, 0.3, 0.4, 0.5]$. 
Furthermore, in our PyTorch implementation of the neural network models, the parameters are initialized the same with~\cite{meng2020incorporating}, the embedding size is $d=100$, the mini-batch size is $512$, and the largest number of epochs is $50$ for fair comparison. Following~\cite{pan2020star}, the normalized weight $w_k$ is set to 12 on three datasets.

\subsection{Overall Performances}


We first compare the top-$K$ recommendation performance with other state-of-the-art methods and set $K=[5, 10, 20]$ to evaluate the performance. Tab.~\ref{tab:overall} shows our experimental results which lead to the following findings.

First, our proposed EMBSR method consistently outperforms all the baselines on three datasets, which clearly demonstrates the effectiveness of utilizing not only the sequential patterns but also the explicit (dyadic) relational patterns of user micro-behaviors. 
Compared with those macro-behavior models, EMBSR goes further to take into account the user's micro-behavior operations which have finer granularity and provide a deeper understanding of the user. 
Compared with the other micro-behavior models, EMBSR works as it exploits the pairwise semantics of micro-behaviors by encoding and examining their dyadic relations directly rather than just relying on the sequential patterns.
According to the \emph{Wilcoxon signed-rank test}, the performance improvements brought by our proposed EMBSR over the best performing baselines are statistically significant with the $p$-values $\ll 0.01$ on all datasets.
     
Second, Deep-learning approaches in general achieve much better performances than traditional methods, such as SKNN. 
Despite the recent improvements made to such non-neural models, they are still limited in detecting and exploiting useful patterns in sessions.
Moreover, we can observe that S-POP method completely fails on Trivago, which means that the ground truth item rarely appears in the session. We believe that it is why EMBSR has different improvements on two metrics of the three datasets.
For Trivago, since the ground truth is not included in the session, the proposed model needs more ability to find it, leading to a huge improvement in \textit{H@K}.  For Appliances and Computers, the information of the ground truth is easier to find; our model is able to give it a higher score to improve the ranking performance.

Third, GNN-based models (SRGNN, GCSAN, SGNN-HN, and MKM-SR) in general outperform the RNN-based models (NARM, RIB, HUP) and the attention-based models (STAMP, BERT4Rec), testifying the superior capacity of GNN in modeling session information. SGNN-HN has the second best performance in most cases, which verifies the effectiveness of SGNN on modeling the complex transition relationship of macro-items. In addition, MKM-SR can outperform SRGNN in most cases, showing that incorporating the micro-behavior by RNN can also improve the performance as MKM-SR considers micro-operation based on the same GNN structure.  
Nevertheless, our proposed EMBSR outperforms all those GNN baselines with a large margin, suggesting that two patterns of micro-behavior encoded by EMBSR probably have captured the most useful information. 

\begin{table*}[tb]
    \centering
    \caption{Performances (\%) of Ablation Studies.}
    \label{tab:Ablation}
    \begin{tabular}{l|cccccccccccc}
    \toprule
    \multirow{2}*{Method} &  \multicolumn{4}{c}{JD-Appliances} & \multicolumn{4}{c}{JD-Computers} & \multicolumn{4}{c}{Trivago} \\
    \cmidrule(lr){2-5} \cmidrule(lr){6-9} \cmidrule(lr){10-13} 
    & H@10 & H@10& M@10 & M@20 & H@10 & H@20& M@10 & M@20 & H@10 & H@20 & M@10 & M@20\\
    \midrule  
    EMBSR-NS & 46.73& 59.32& 21.12& 22.00 &32.19& 43.85& 12.51 & 13.32 & \textbf{23.04}&  \textbf{31.74}& 9.76& 10.36\\
    EMBSR-NG & 46.78&  59.43 &  19.04 &  19.92&  32.04&  44.02&  11.91&  12.74 & 22.19&  30.52& 9.96&10.53\\
    EMBSR-NF & 48.99& 60.79& \textbf{25.60}& \textbf{26.42}&  15.17& 33.35& 44.66&  15.96&  9.90& 20.56& 26.80& 10.33 \\
	EMBSR & \textbf{49.57}&\textbf{61.64}&  25.21& 26.07& \textbf{34.75}& \textbf{46.29} & \textbf{15.38}& \textbf{ 16.18} & 22.95&  31.18& \textbf{10.00}&\textbf{10.56} \\
	\bottomrule
    \end{tabular}
\end{table*}
\begin{figure*}[tb]
	\centering
    \includegraphics[width= 0.85\textwidth]{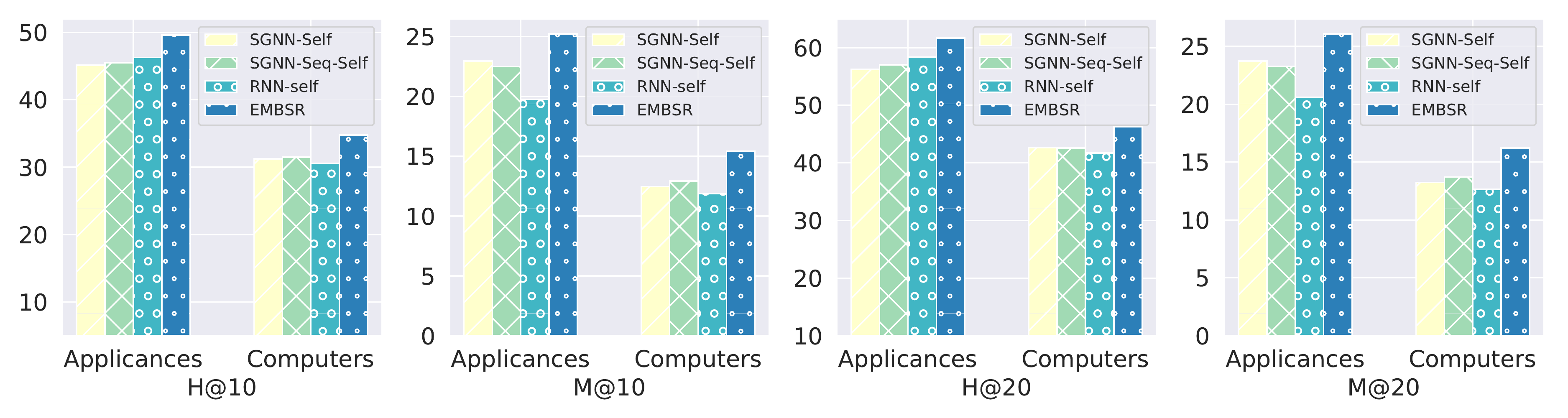}
	\caption{Performance (\%) comparison to assess the utility of the sequential pattern of micro-behaviors.}
	\label{fig:seq_beh}
\end{figure*}
\begin{figure*}[t]
	\centering
    \includegraphics[width= 0.85\textwidth]{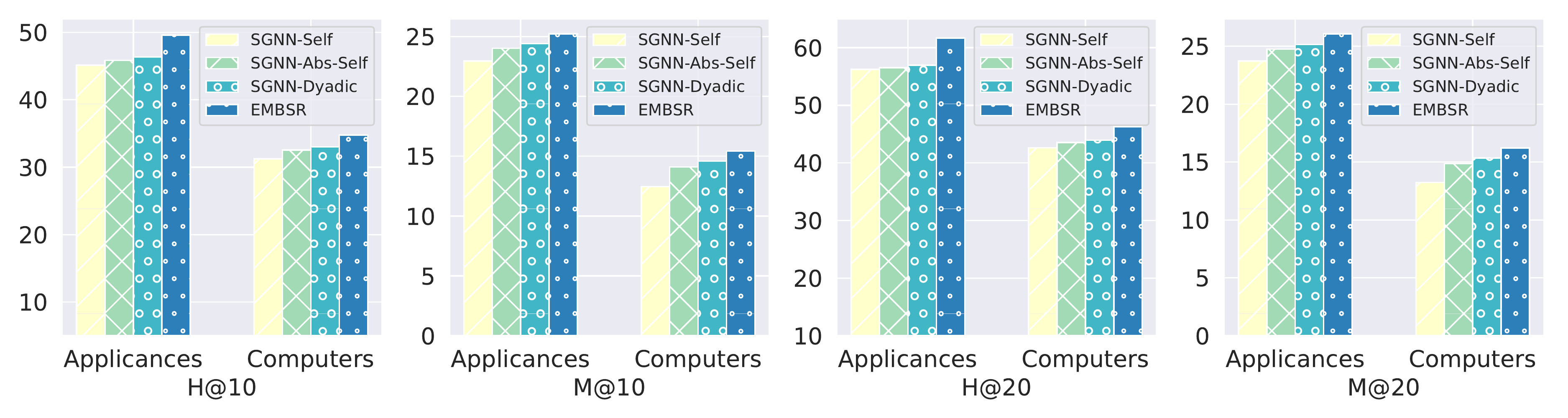}
	\caption{Performance (\%) comparison to assess the utility of the dyadic relation pattern of micro-behaviors.}
	\label{fig:dyadic_beh}
\end{figure*}
\subsection{Ablation Studies}
In order to measure the contribution from the different components of the proposed model, we contrast it with three ablated versions and set $K=[10, 20]$ to evaluate the performance: 
\begin{itemize}
    \item \textbf{EMBSR-NS} removes the operation-aware self-attention layer and only encodes the sequential pattern of micro-behaviors for the task. 
    \item \textbf{EMBSR-NG} removes the entire GNN Layer, including the GRU layer for micro-operation sequence, and only encoding the dyadic relational pattern of micro-behaviors for the task.
    \item \textbf{EMBSR-NF} removes the fusion gate network for the final representation, while we directly concatenate these two embeddings and feed it into an MLP for the representation of the session
\end{itemize}

Tab.~\ref{tab:Ablation} presents the results of the ablation studies for three datasets.
In Appliances and Computers, EMBSR-NS and EMBSR-NG yield the worst performance, confirming that only modeling a single pattern of the session cannot fully capture the user's real intent and preference; In comparison, EMBSR-NS is in general slightly better than EMBSR-NG, especially on $M@K$, proving that the GNN-based model incorporating the sequential patter of micro-behaviors has indeed helped to make better predictions;
EMBSR-NF in general has the second best performance, which has further demonstrated the effectiveness of explicitly modeling two different patterns of micro-behaviors since it only uses another way to generate the representation of the session by these two patterns.

In Trivago, the results are slightly more complicated. EMBSR-NF generally yields the worst performance, while EMBSR-NS and EMBSR-NR both deliver the competitive performance, indicating that the correct fusion mechanism is also essential in the dataset. Therefore, the fact that the full EMBSR model in general performs best confirms the advantages of fusing different representations by the fusion gating network.

\begin{figure*}[tb]
	\centering
	\subfigure[JD-Appliances]{
    \includegraphics[width= 0.85 \textwidth]{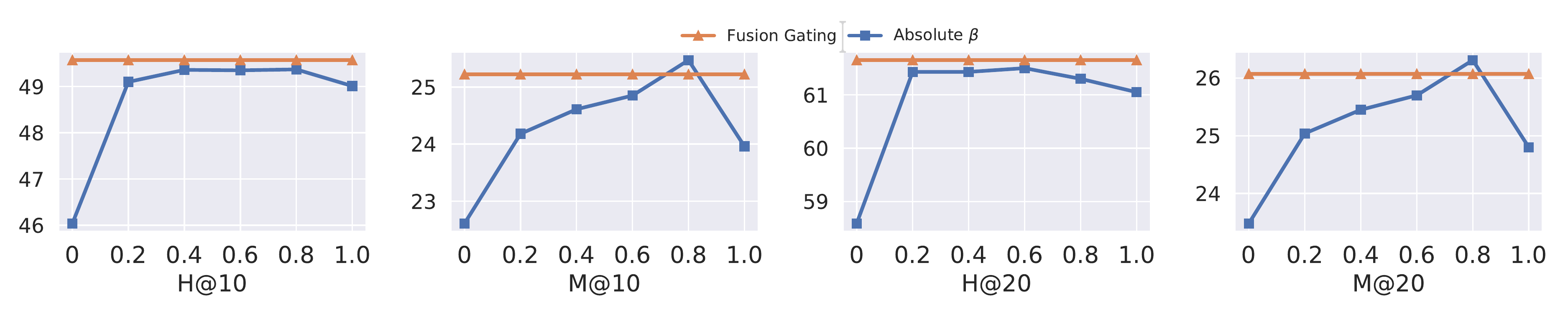}
}
	\subfigure[JD-Computers]{
	\includegraphics[width= 0.85\textwidth]{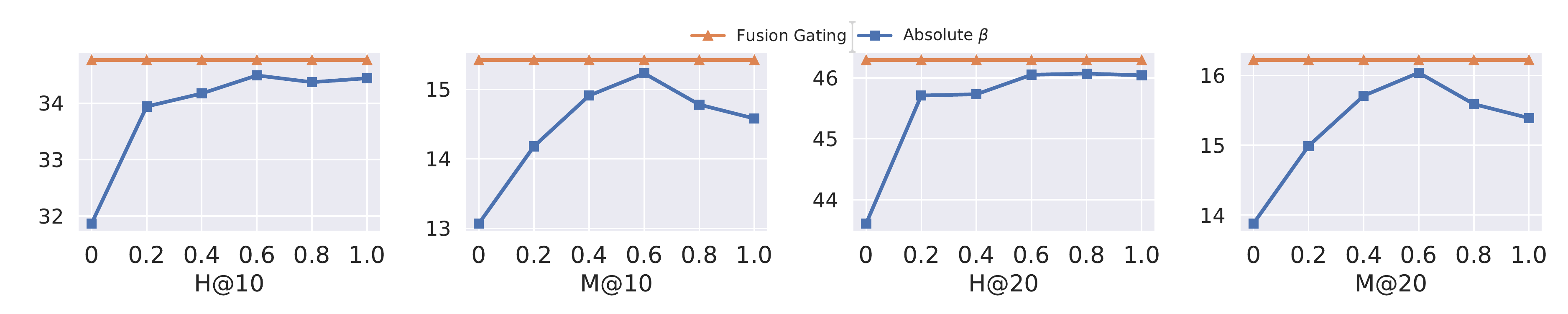}
}
	\caption{Performance (\%) comparison to assess the utility of the fusion gating mechanism.}
	\label{fig:beta}
\end{figure*}

\subsection{Utility of Sequential Patterns}

To understand the impact of the sequential pattern, we choose the Appliances and Computers datasets for further investigation since they have more types of micro-operations. Since the motivation of this paper is to utilize the micro-behaviors to help improve the recommendation quality on items, we  also design three variants to investigate the way of incorporating the different patterns of micro-behaviors:
\begin{itemize}
    \item \textbf{SGNN-Self} uses the star graph without GRU and a standard self-attention mechanism. It has no information of micro-behaviors and can only learn the representation of the session by macro-items.
    \item \textbf{SGNN-Seq-Self} encodes the information of sequential pattern on the SGNN with GRU based on SGNN-Self.
    \item \textbf{RNN-Self} replaces the whole GNN layer with an RNN layer compared with SGNN-Self. It directly concatenates the item embedding and operation embedding of the initial session and feeds them to the RNN for learning the sequential pattern of micro-behaviors. 
\end{itemize} 

Fig.~\ref{fig:seq_beh} illustrates the result of the comparison. We see that EMBSR achieves the best performance in all cases. As for the variants, SGNN-Seq-Self in general performs better than SGNN-Self. We attribute this to the sequential pattern of micro-behaviors, which are related to users’ preference of the macro-items. However, the RNN-based method, RNN-Self, performs worst in most cases, especially on $M@10$ and $M@20$. Although it also includes the micro-operations information, only a simple encoding method based on RNN cannot capture the complex transition pattern of macro-items and make full use of micro-behavior information. Thus, EMBSR presents an approach to encode the sequential pattern of micro-behaviors in GNN, including a novel method to convert a session into a multigraph, which is able to model the user's real intention and preference.

\subsection{Utility of Dyadic Relational Patterns}

To demonstrate the impact of dyadic encoding, we compare the experimental results with two other variants: 
\begin{itemize}
    \item \textbf{SGNN-Abs-Self} replaces operation-aware self-attention and dyadic encoding with standard self-attention and absolute operation embedding.
    \item \textbf{SGNN-Dyadic} encodes dyadic relational patterns but only uses SGNN for macro-items without RNN.
\end{itemize}

We also add SGNN-Self and EMBSR for a clear comparison. As illustrated in Fig.~\ref{fig:dyadic_beh}, we observe that RNN-Self has achieved the worst performance again since it does not have any micro-behavior information. Moreover, SGNN-Dyadic outperforms SGNN-Abs-Self in all cases. Here, SGNN-Abs-Self encodes the micro-behaviors by a simple way that adds the absolute operation embedding in a standard self-attention network. It cannot capture pair-wise semantic of the micro-operations well, which is important to understand the difference between two behaviors. Furthermore, there are some surprising improvements about the performance of SGNN-Dyadic, which is lack of sequential patterns of micro-behaviors, but still achieves competitive results, especially on $M@K$ of Computers. This indicates that the dyadic relational pattern, which has never been exploited or discussed in SR, is essential to model the user's real preference on the item.   
\begin{figure*}
	\centering
 	\includegraphics[width=0.8\textwidth]{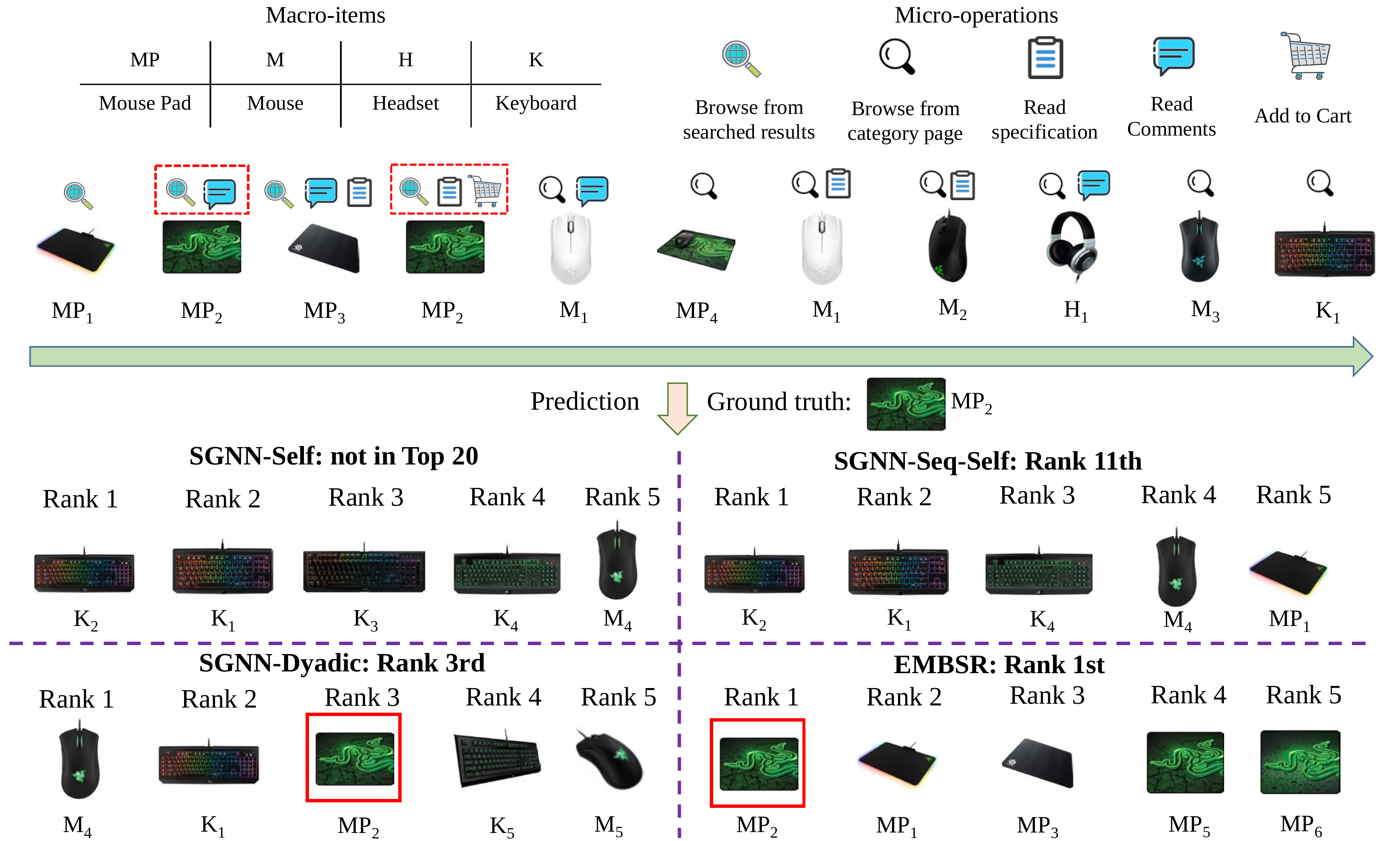}
 	\caption{A case for the session-based recommendation with the micro-behavior.}
 	\label{fig:case_study}
\end{figure*}
\subsection{Utility of Fusion Gating Mechanism}

In order to investigate the utility of the fusion gating mechanism in equation~\ref{fusion}, we tune $\beta$ in $ \{0, 0.2, 0.4, 0.6, 0.8, 1\}$ and use it as an absolute weight to control the fusion of different information. Specifically, $x_t$ represents the recent interest of the user, but $z_s$ has both sequential and dyadic encoding information. $\beta$ controls what information is used to generate the final representation.

As illustrated in Fig. \ref{fig:beta}, it is reasonable to infer that how to select the information is crucial to the final performance. When $\beta = 0$, we only use the recent interest $x_t$, and it achieves the worst performance, evidencing that only the recent interest cannot fully express the user's preference. When $\beta = 1$, the result is competitive since $z_s$ has considered all information of micro-behaviors within the session. However, due to too much information, it may not be able to capture the user's recent interest. Thus, a simple way for the information fusion with an absolute $\beta$ can achieve better performance. Moreover, the final result is not always sensitive to $\beta$, since the learning process can automatically adapt to achieve a better result. The proposed fusion gating network avoids the choice of $\beta$, and has the best results in general.

\subsection{Case Study}

Fig.~\ref{fig:case_study} illustrates a real case from "Computers" dataset to demonstrate the rationality of our proposed framework. This session has a total of 20 micro-behaviors on 11 macro items. The bottom part is the prediction results of three variants (SGNN-Self, SGNN-Seq-Self, SGNN-Dyadic) and EMBSR for this session. We also list the top five items that each method recall and can observe that:

(1) The macro-item sequence only reflect the coarse-grained preference of the user. From the item sequence of the case, we can infer that the user tends to pick out one computer accessory. In general, the recent items have more impacts on the next click~\cite{liu2018stamp}. Therefore, the top five items recommended by the SGNN-Self are all keyboards since the last item is a keyboard. However, the keyboard is far from the user's real intention, resulting in the failed recall of SGNN-Self.

(2) Micro-behaviors play a significant role in understanding users' fine-grained preferences for items. In this case, the user has obvious signals for the mouse pad. For MP$_2$ and MP$_3$, the user reads the detail specification and comments after click, and finally adds MP$_2$ to the shopping cart. Therefore, SGNN-Seq-Self, SGNN-Dyadic, and EMBSR successfully recall the MP$_2$ in the top 20 since they all have considered the micro-operation signals. Moreover, EMBSR integrates sequential patterns with dyadic relational patterns, which enables the model to capture the real intention of the user. As we can see, the top five items recalled by EMBSR are all mouse pads. Noting that MP$_2$, MP$_5$, and MP$_6$ are the same items with different sizes (medium, small and large, respectively), which proves that EMBSR understands the user's preferences and accurately captures the similarities among items. 

\section{Conclusions}

In this paper, we propose a novel approach to SR --- EMBSR --- which considers not only the sequential patterns but also the dyadic relational patterns of micro-behaviors within each session. 
Specifically, we have designed a graph neural network to aggregate the sequence of micro-behaviors, and developed an operation-aware self-attention mechanism to extract the pair-wise semantics of micro-behaviors. 
Our experiments have demonstrated that the proposed EMBSR model significantly outperforms all state-of-the-art SR methods. 
For future work, it would be interesting to investigate how to exploit other operations that are for all items such as filtering and sorting to further improve the performance of SR systems, and whether it would be beneficial to weight, or filter, micro-behavior operations according to their importance.

\section*{Acknowledgment}
This work was supported by NSFC grants (No. 62136002 and 61972155), National Key R\&D Program of China (No. 2021YFC3340702), the Science and Technology Commission of Shanghai Municipality (20DZ1100300) and the Open Project Fund from Shenzhen Institute of Artificial Intelligence and Robotics for Society, under Grant No. AC01202005020, Shanghai Trusted Industry Internet Software Collaborative Innovation Center.

\bibliographystyle{IEEEtranN}
\bibliography{reference}
\vspace{12pt}


\clearpage

\begin{center}
\textbf{\large Supplemental Materials}
\end{center}
\setcounter{equation}{0}
\setcounter{figure}{0}
\setcounter{table}{0}
\setcounter{page}{1}
\makeatletter


\setcounter{section}{0}

\section{Additional experiments}

\subsection{Optimal performance for macro-behavior baselines}

Since most of existing methods of session-based recommendation are designed for the sequence of macro item, the performance of macro-behavior baselines utilizing micro-behavior may not be able to reach their optimal performance if there are a large number of operations which are treated equally. Therefore, we have identified one type of operation and used that to redefine the sequence of items for the additional experiment. Since most methods for SR use the clickstream data, we only keep the click-related data in two JD datasets and "click-outs" data in Trivago for macro-behavior baselines. Meanwhile, we add other operations to the sequence while ensuring that the ground truth of each sequence is consistent for a fair comparison with our proposed approach. 

The result is reported in Tab.~\ref{tab:o4}, we compare EMBSR with the best baseline SGNN-HN and the typical baseline BERT4Rec. This table shows similar patterns with using all operations for macro-behavior models. In addition, we have achieved a considerable improvement on Trivago. Intuitively, EMBSR that are able to exploit the information embedded in all the operations does work better than those that are restricted to utilize only some of the operations. Macro-behavior models do have many limitations. 

Furthermore, assigning different importance weights to different operations is probably more a promising way than simply discarding some insignificant operations. It is indeed an interesting question whether it would be beneficial to weight, or filter, micro-behavior operations according to their importance. Besides, the importance of micro-behavior operations may be not static but vary in different sequences or at different positions. This line of thoughts, though interesting, would make the model much more complicated than its current form. We have to leave the bulk of experiments for the future work.

\begin{table}[tb]
    \centering
    \caption{Performances (\%) of defining the sequence of items with the single type of operation.}
    \label{tab:o4}
   {
    \begin{tabular}{l|l|cccc}
    \toprule
    Datasets& Metrics &BERT4Rec &SGNN-HN& EBMSR & Imp \\
    \midrule  
    \multirow{6}*{Appliances} 
    &H@5& 30.16 & \underline{34.95} & \textbf{37.41} & 7.04\%\\
    &H@10& 41.50 & \underline{47.17} & \textbf{49.76} & 5.49\%\\
    &H@20& 52.97 & \underline{59.56} & \textbf{62.00}& 4.97\%\\
    &M@5& 16.54 & \underline{21.34} & \textbf{23.74}& 11.25\%\\
    &M@10& 18.05 & \underline{22.96} & \textbf{25.40}& 10.63\%\\
    &M@20& 18.85 & \underline{23.82} & \textbf{26.25}& 10.20\%\\
    
    \hline
    \multirow{6}*{Computers} 
    &H@5& 17.54 & \underline{21.34} & \textbf{24.23} & 13.54\%\\
    &H@10& 26.67 & \underline{31.93} & \textbf{34.91}& 9.33\%\\
    &H@20& 37.11 & \underline{43.72} & \textbf{46.50}& 6.36\%\\
    &M@5& 8.81 & \underline{11.49} & \textbf{13.83}& 20.37\%\\
    &M@10& 10.02 & \underline{12.89} & \textbf{15.24}&18.23\% \\
    &M@20& 10.74 & \underline{13.71} & \textbf{16.04}&16.99\% \\
    \hline
    \multirow{6}*{Trivago} 
    &H@5& 13.98 & \underline{17.21} & \textbf{25.03} & 45.44\%\\
    &H@10& 17.51 & \underline{22.69} & \textbf{29.89}&31.73\% \\
    &H@20& 20.48 & \underline{27.94} & \textbf{34.92}&24.98\% \\
    &M@5& 8.57 & \underline{10.72} & \textbf{18.97}& 76.96\%\\
    &M@10& 9.04 & \underline{11.44} & \textbf{19.61}&71.42\% \\
    &M@20& 9.25 & \underline{11.81} & \textbf{19.96}&69.01\%\\
	\bottomrule
    \end{tabular}
    }
\end{table}

\subsection{Dyadic relational encoding with SGNN-HN}
In Section V-E (Figure 5), we do witness the high utility of dyadic relational patterns for recommendation effectiveness. In fact, \textbf{SGNN-Dyadic} is the model that we isolate the idea of dyadic encoding to the best macro-behavior baseline SGNN-HN, which has achieved the competitive results. We have reported their performances on two JD datasets in Tab.~\ref{tab:o5} here. From this table, we observe that EMBSR-Dyadic outperforms SGNN-HN, except for H@10 and H@20 on Appliances, further confirming the importance of the dyadic relationship and the generalizability of the proposed idea. In addition, EMBSR still achieves a large improvement, showing that the proposed encoding scheme that transforms a session into a directed multigraph and the novel aggregation stage are more suitable for micro-behavior encoding.

\begin{table}[tb]
    \centering
    \caption{Performances (\%) of applying the dyadic encoding to SGNN-HN.}
    \label{tab:o5}
   {
    \begin{tabular}{l|l|ccc}
    \toprule
    Datasets& Metrics &SGNN-HN &EMBSR-Dyadic& EBMSR \\
    \midrule  
    \multirow{6}*{Appliances} 
    &H@5& 34.80 & \underline{35.64} & \textbf{37.34} \\
    &H@10& \underline{47.04} & 46.36 & \textbf{49.57}\\
    &H@20& \underline{59.36} & 56.94 & \textbf{61.64}\\
    &M@5& 21.00 & \underline{22.98} & \textbf{23.58}\\
    &M@10& 22.64 & \underline{24.41} & \textbf{25.21}\\
    &M@20& 23.49& \underline{25.15} & \textbf{26.06}\\
    
    \hline
    \multirow{6}*{Computers} 
    &H@5& 21.53 & \underline{23.09} & \textbf{24.17}\\
    &H@10& 32.01 & \underline{32.99} & \textbf{34.75}\\
    &H@20& 43.67 & \underline{43.92} & \textbf{46.29}\\
    &M@5& 11.61 & \underline{13.28} & \textbf{13.98}\\
    &M@10& 13.00 & \underline{14.59} & \textbf{15.38}\\
    &M@20& 13.81 & \underline{15.35} & \textbf{16.18}\\
	\bottomrule
    \end{tabular}
    }
\end{table}
\subsection{Top Ranked Results}
\begin{table*}[htbp]
    \centering
    \caption{Performances (\%) of $K=[1,3,5]$. The highest scores are boldfaced; the 2nd highest scores are underlined.}
    \label{tab:top1}
    \resizebox{\textwidth}{!}{
    \begin{tabular}{l|l|cccccccc|ccc|cc}
    \toprule
    Datasets& Metrics &S-POP & SKNN& NARM & STAMP & SR-GNN & GC-SAN& BERT4Rec & SGNN-HN & RIB & HUP & MKM-SR & EMBSR & Imp.\\
    \midrule  
    \multirow{6}*{Appliances} 
     &H@1&9.38&6.97&10.84&11.39&12.46&11.05&9.15&13.48&9.65&10.02&\underline{13.49}&\textbf{16.06}& 19.05\%\\
     &H@3&23.45&17.49&23.20&23.41&25.13&22.91&23.07&\underline{26.67}&22.73&23.93&26.32&\textbf{29.24}&9.64\%  \\
     &H@5&31.66&25.06&30.94&30.74&32.65&30.36&31.02&\underline{34.80}&30.12&31.91&33.82&\textbf{37.34}& 7.30\% \\
     &M@1&9.38&6.97&10.84&11.39&12.46&11.05&9.15&13.48&9.65&10.02&\underline{13.49}&\textbf{16.06}& 19.05\%\\
     &M@3&15.42&11.42&16.14&16.54&17.92&16.14&15.15&\underline{19.15}&15.29&16.02&19.03&\textbf{21.74}& 13.52\%\\
     &M@5&17.29&13.15&17.90&18.21&19.63&17.83&16.96&\underline{21.00}&16.97&17.83&20.73&\textbf{23.58}& 12.29\% \\
    
    \hline
    \multirow{6}*{Computers} 
     &H@1&5.27&4.21&4.79&5.86&6.80&4.32&4.95&6.46&5.17&5.52&\underline{7.24}&\textbf{8.64}& 19.34\% \\
     &H@3&12.60&10.35&12.62&13.07&14.66&12.72&12.58&15.22&12.24&13.54&\underline{15.43}&\textbf{17.78}& 15.23\% \\
     &H@5&17.18&15.11&18.31&18.18&20.08&18.79&17.90&\underline{21.53}&16.93&18.87&21.00&\textbf{24.17}& 12.26\%\\
     &M@1&5.27&4.21&4.79&5.86&6.80&4.32&4.95&6.46&5.17&5.52&\underline{7.24}&\textbf{8.64}& 19.34\% \\
     &M@3&8.41&6.82&8.11&8.94&10.15&7.88&8.21&10.18&8.19&8.94&\underline{10.75}&\textbf{12.54}& 16.65\% \\
     &M@5&9.45&7.89&9.40&10.09&11.38&9.26&9.42&11.61&9.26&10.15&\underline{12.01}&\textbf{13.98}& 16.40\% \\
          \hline
    \multirow{6}*{Trivago} 
     &H@1&0&0.05&4.68&5.31&4.87&4.39&4.20&\textbf{5.64}&3.80&3.81&4.95&\underline{5.49}& -2.66\% \\
     &H@3&0&4.34&9.73&10.19&9.43&10.19&8.37&\underline{11.14}&7.20&7.61&9.57&\textbf{11.62}&4.31\% \\
     &H@5&0&7.89&12.89&13.11&11.97&14.15&11.01&\underline{14.58}&9.00&10.06&12.34&\textbf{15.80}& 8.37\%\\
     &M@1&0&0.05&4.68&5.31&4.87&4.39&4.20&\textbf{5.64}&3.80&3.81&4.95&\underline{5.49}& -2.66\% \\
     &M@3&0&1.85&6.85&7.42&6.84&6.85&6.00&\underline{8.01}&5.28&5.44&6.94&\textbf{8.10}& 1.12\%\\
     &M@5&0&2.65&7.57&8.09&7.42&7.76&6.60&\underline{8.79}&5.69&6.00&7.58&\textbf{9.05}& 2.96\% \\ 
	\bottomrule
    \end{tabular}
    }
\end{table*}
In order to more comprehensively evaluate the performance of our proposed EMBSR, we have reported the result of top 1,3,5 in Tab.~\ref{tab:top1}. From this table, we can observe that the performances of top 1, 3, and 5 results exhibit similar patterns with top 10, and 20. It is worth noting that there is no difference between \emph{H@1} and \emph{M@1}, so they have the same values. In addition, as we analyzed in the paper, since the ground truth is not included in the input session, our proposed approach has not achieved the best performance on Trivago when $K=1$. 
\end{document}